\documentclass[12pt]{article}
\usepackage{aaspp4}

\begin{document}

\title{On the normalization of the QSO's Ly$\alpha$ forest power spectrum}

\author{Priya Jamkhedkar\footnote{priya@physics.arizona.edu},
  Hongguang Bi\footnote{bihg@time.physics.arizona.edu} and 
 Li-Zhi Fang\footnote{fanglz@physics.arizona.edu}}

\affil{Department of Physics, University of Arizona, Tucson, AZ
85721}

\begin{abstract}

The calculation of the transmission power spectrum of QSO's Ly$\alpha$ 
absorption requires two parameters for the normalization: the continuum 
$F_c$ and mean transmission $\overline{e^{-\tau}}$. Traditionally, the 
continuum is obtained by a polynomial fitting truncating it at a lower 
order, and the mean 
transmission is calculated over the entire wavelength range considered. The 
flux $F$ is then normalized by $F_c\overline{e^{-\tau}}$. However, the 
fluctuations in the transmitted flux are significantly correlated with the 
local background flux on scales for which the field is intermittent. 
As a consequence, the normalization of the entire power spectrum by an 
over-all mean transmission $\overline{e^{-\tau}}$ will overlook the 
effect of the fluctuation-background correlation upon the powers.  
In this paper,  we develop a self-normalization algorithm of the 
transmission power spectrum based on a multiresolution analysis. 
This self-normalized power spectrum estimator needs neither a continuum
fitting, nor pre-determining the mean transmission. With simulated 
samples, we show that the self-normalization algorithm can perfectly 
recover the transmission power spectrum from the flux regardless of how 
the continuum varies with wavelength. We also show that the 
self-normalized power spectrum is also properly normalized by the
mean transmission. Moreover, this power spectrum estimator is sensitive
to the non-linear behavior of the field. That is, the self-normalized 
power spectrum estimator can distinguish between fields with or without the 
fluctuation-background correlation. This cannot be accomplished by the 
power spectrum with the normalization by an overall mean transmission. 
Applying this 
analysis to a real data set of q1700+642 Ly$\alpha$ forest, we demonstrate 
that the proposed power spectrum estimator can perform correct normalization,
and effectively reveal the correlation between the fluctuations and 
background of the transmitted flux on small scales. Therefore, the 
self-normalized power spectrum would be useful for the discrimination 
among models without the uncertainties caused by free (or fitting) 
parameters. 

\end{abstract}


\section{Introduction}

Ly$\alpha$ absorption, shortward of Ly$\alpha$ emission in QSO spectra, 
indicates the presence of intervening absorbers with neutral hydrogen 
column densities  ranging from about $10^{13}$ to $10^{22}$ cm$^{-2}$. The 
absorbers with low column densities, {\it e.g.} from $10^{13}$ to 
$10^{17}$ cm$^{-2}$, are usually called Ly$\alpha$ forest. It is 
generally thought that the low column density absorbers are some kind 
of weakly clustered clouds consisting of photoionized intergalactic gas 
({\it e.g.} Wolfe 1991; Bajtlik 1992). This suggests that Ly$\alpha$ 
forests are caused by diffusely 
distributed IGM in pre-collapsed areas of the cosmic mass field
(Bi, 1993; Fang et al. 1993; Bi, Ge \& Fang 1995; Bi \& Davidsen 
1997.) Observations of the size and velocity dispersion of the Ly$\alpha$ 
clouds at high redshift also show that the absorption probably is not 
caused by confined objects at high redshifts (Bechtold et al. 1994; 
Dinshaw et al. 1994; Fang et al 1996; Crotts \& Fang 1998.) 
With this picture, the baryonic matter distribution is almost 
point-by-point proportional to the dark matter distribution on all 
scales larger than the IGM's Jeans length, i.e.  the Ly$\alpha$ forests 
would be good tracers of the underlying dark matter distribution.

Thus, the power spectrum of QSO Ly$\alpha$ transmitted flux can be used to 
estimate the power spectrum of the underlying mass field, and then be used
to constrain cosmological parameters (Croft et al 1999;  McDonald et al 
1999; Hui 1999; Feng \& Fang 2000.) A key step in this approach is
to compare the power spectrum of observed transmitted flux fluctuations
with model-predicted power spectrum. One uncertainty in the power 
spectrum determination of the real data is from the normalization 
of the power spectrum.  
Therefore, in order to have an effective confrontation between the
observed and theoretical power spectrum of Ly$\alpha$ forests, 
it is necessary to develop a proper algorithm for the normalization of  
the power spectrum. This is the goal of this paper.

The observed flux of a QSO absorption spectrum is given by 
$F(\lambda)= F_c(\lambda)e^{-\tau}$, where $F_c(\lambda)$ is the 
continuum, $e^{-\tau(\lambda)}$ the transmission, and 
$\tau$ the optical depth. The normalized power spectrum of 
transmission is the power spectrum of the transmission flux 
fluctuations $\delta(\lambda)$, defined as
%eq1
\begin{equation}
\delta(\lambda) = \frac{F(\lambda)- \overline{F(\lambda)}}
  {\overline{F(\lambda)}}.
\end{equation}
That is, the transmission power spectrum is normalized by the mean 
flux 
$\overline{F(\lambda)}=F_c(\lambda) \overline {e^{\tau(\lambda)}}$.
In other words, the normalization of the transmission power spectrum 
is determined by two factors: the continuum $F_c(\lambda)$ and the mean 
transmission $\overline {e^{\tau(\lambda)}}$.

Traditionally, the continuum is needed to be determined before the 
power spectrum calculation. Usually the continuum is obtained by 
a fitting of polynomial or its variants. Assuming that the continuum 
fluctuates slowly, the polynomial or its variants are truncated at 
relatively low orders (e.g. Croft et al 2000; Hui et al 2000). The 
pre-assumed polynomial or other function, and the subsequent truncation 
may lead to uncertainty of the power spectrum.

Another source of uncertainty of the transmission power spectrum
is the mean transmission $\overline{e^{-\tau}}$ normalization.
The mean transmission is calculated by averaging the flux over the 
entire wavelength range considered, and the power spectrum is 
normalized by this mean transmission for all scales. This implicitly 
assumes that there is no correlation between the transmitted flux 
fluctuations and the mean flux. This assumption is true for a gaussian 
field, but may not be so for a non-linearly evolved field. 

In fact, the fluctuations at position $\lambda$ are correlated with 
the background at the same position $\lambda$. Recent findings that 
the transmitted flux of Ly$\alpha$ forests exhibits intermittent behavior 
(Jamkhedkar, Zhan \& Fang 2000) clarifies this point. 
That is, the transmitted flux shows prominent spiky feature fluctuations 
on small scales. The transmitted flux consists of rare but strong density 
fluctuations randomly scattered in space with very low fluctuations in 
between. In this case, the power of the transmission fluctuations is
mainly dominated by the spikes. On the other hand, the transmission 
$e^{-\tau}$ is low at the spikes. That is, the transmission fluctuations 
are anti-correlated with transmission. As a consequence, the power 
would be underestimated if the power spectrum is normalized 
by the mean transmission $\overline{e^{-\tau}}$ over the entire wavelength 
range. Since the spiky features are stronger on smaller scales, the 
normalization by an over-all mean transmission $\overline{e^{-\tau}}$
or by a filling factor with a scale-independent flux threshold 
(Croft et al 2000) will cause an underestimation of power on 
small scales.

Recently, we have developed a power spectrum estimator with a 
multiresolution analysis based on the Discrete Wavelet Transform (DWT).  
The DWT power spectrum estimator is found to be very useful for the 
recovery of the initially linear power spectrum (Feng \& Fang 
2000;  Pando, Feng \& Fang, 2001). In this paper, we show that 
the DWT algorithm is also very useful to detect the power spectrum 
of non-linear field, like an intermittent field. The DWT 
algorithm can effectively reduce the above-mentioned uncertainties 
due to free parameters used for normalization. We will 
show that the normalization of a  DWT power spectrum does not rely 
on a continuum fitting and the mean transmission. Moreover, the 
power spectrum given by this estimator is sensitive to the 
correlation between the flux fluctuations and the background flux. 
That is, the power spectrum can be employed to distinguish
among the fields with and without intermittency. Therefore, it would
be useful for discrimination among models of the Ly$\alpha$ forests.  

The paper will be organized as follows. \S 2 introduces briefly the  
Discrete Wavelet Transform (DWT) analysis of the flux of QSO 
absorption spectrum. \S 3 presents the self-normalization algorithm 
of the transmission power spectrum. It doesn't need either a continuum 
fitting,  or a calculation of the mean transmission. In \S 4,
we test the self normalization algorithm. We show that the 
self-normalization algorithm can effectively perform the normalization
due to either continuum or mean transmission.  \S 5 will demonstrate 
that the DWT self-normalized power spectrum is useful to detect 
the intermittent behavior of the field.  Finally, the conclusions and 
discussions will be presented in \S 6.

\section{The DWT analysis of the QSO Ly$\alpha$ forest flux}

\subsection{Need for a space-scale decomposition}

We rewrite eq.(1) as 
%eq2
\begin{equation}
F(\lambda) = \overline{F(\lambda)} 
   +\overline{F(\lambda)} \delta(\lambda).
\end{equation}
Our purpose is to estimate the power spectrum of the flux fluctuations
$\delta(\lambda)$ from the observed flux $F(\lambda)$. Therefore, eq.(1) 
requires one to decompose the observed flux $F(\lambda)$ into two terms: the 
first one is the background $\overline{F(\lambda)}$, which does not contain 
information of the fluctuations considered, while the second term 
contains all this information. If the background $\overline{F(\lambda)}$ 
is $\lambda$ dependent and correlated with the fluctuations 
$\delta(\lambda)$, the decomposition eq.(1) apparently cannot be 
done by truncating at {\it a priori} ``relative low orders".

We will solve this problem by a scale-by-scale analysis, without introducing
new parameters. In terms of a scale-by-scale analysis, eq.(1) means that 
to detect the power of the flux fluctuations on the scale $r$,  all  
components of $F(\lambda)$ on scales larger than $r$ play the role of a 
background. Therefore, to determine power on the scale $r$, one can
decompose the observed $F(\lambda)$ into two terms: the first doesn't 
contain any information on scales larger than $r$, while the second 
contains all information on scales equal to or less than $r$. 

This decomposition refers to both the position $\lambda$ and the scale 
$r$, and therefore, we need a scale-space decomposition of the 
transmitted flux.

On the other hand, the calculation of power spectrum essentially is a 
decomposition of the flux into scale domains. Therefore, it is possible 
to do the decomposition of eq.(2) by the same scale-space decomposition as 
that used for measuring the power spectrum. In other words, the estimation of 
the normalization background and the calculation of power spectrum can 
be accomplished simultaneously. Once the orthonormal bases for power 
spectrum estimation are given, the term $\overline{F(\lambda)}$ is 
uniquely determined without free (or fitting) parameters.

The Fourier power spectrum is not convenient for this purpose, as the 
bases of Fourier transform are not localized in physical space, and they
don't yield a  scale-space decomposition. We will use the DWT, whose bases 
are localized both in scale and position space (e.g. Mallat, 
1989a,b,c; Meyer, 1992; Daubechies, 1992; and references therein, and 
for physical applications, refer to Fang \& Thews, 1998.) 

\subsection{Expansion by scaling functions}

A sample of QSO Ly$\alpha$ absorption spectrum gives a list of the flux 
$F(\lambda)$ observed at discrete wavelength $\lambda_i$, $i=1,2... n$. 
Since $\lambda_i$ corresponds to spatial position $x_i$, or redshift $z_i$, 
one can define a spatial distribution of the flux by
%eq3
\begin{equation}
F(x)= \sum_{i=1}^{n-1}\frac{1}{2}[F(\lambda_i)+F(\lambda_{i+1})]
[\theta(x -\lambda_i) - \theta(x-\lambda_{i+1})], 
\end{equation} 
where the step function $\theta(x)$ is equal to 1, for $x \geq 0$, and 0
for $x<0$. The spatial range $L$ corresponds to the wavelengths from 
$\lambda_{min}$  to $\lambda_{max}$.

In the DWT analysis, the space $L$ is chopped into $2^j$ segments 
labelled by $l=0,1,...2^j-1$. Each of these segments has a size of
$L/2^j$. 
The index $j$ is an integer. Therefore, the index $j$ stands for 
scale $L/2^j$, 
while $l$ is for position, or the spatial range $lL/2^j < x < (l+1)L/2^j$.

We first introduce the scaling functions for the Haar wavelets. They 
are the top-hat window functions defined by 
%eq4
\begin{equation}
\phi_{j,l}^{H}(x) =\left (\frac{2^j}{L}\right)^{1/2}
   \left\{ \begin{array}{ll} 
   1 & \mbox{for $Ll2^{-j} \leq x \leq L(l + 1)2^{-j}$,}\\ 
   0 & \mbox{otherwise.}
\end{array} \right. 
\end{equation}
where the superscript $H$ stands for Haar, and the factor $\sqrt{2^j/L}$
ensures normalization, i.e.
%eq5
\begin{equation}
\int_0^{L} \phi_{j,l}^{H}(x)\phi_{j,l'}^{H}(x)dx= \delta^K_{ll'},
\end{equation}
where $\delta^K$ is Kronecker delta function. 

The scaling function, $\phi_{j,l}^{H}(x)$ is actually a window on 
a resolution scale $L/2^j$ at the position   
$Ll2^{-j} \leq x \leq L(l + 1) 2^{-j}$. But it is not normalized
like a window function $W(x)$ which satisfies
%eq6
\begin{equation}
\int W(x)dx=1.
\end{equation}
Nevertheless
the mean flux in the spatial range  $Ll2^{-j} \leq x \leq L(l + 1) 2^{-j}$ 
is proportional to 
%eq7
\begin{equation}
\epsilon^F_{j,l}= \int_{0}^{L}F(x)\phi_{j,l}^{H}(x)dx.
\end{equation}
The number $\epsilon^F_{j,l}$ is called the scaling function 
coefficient (SFC). 

Using SFCs, one can construct the  flux as
%eq8
\begin{equation}
F^{j}(x) = \sum_{l=0}^{2^{j}-1}\epsilon^F_{j,l}
\phi_{j,l}^{H}(x).
\end{equation}
$F^{j}(x)$ is the flux $F(x)$ smoothed on scale $L/2^j$ (or simply the  
scale $j$). A higher value for $j$ corresponds to smaller scales 
and vice versa. For a given sample with resolution $\delta\lambda$, 
the original flux can be expressed as $F(x)=F^{J}$, where
$J$ is given by the integer of number 
$\log_2 [(\lambda_{max}-\lambda_{min})/\delta \lambda]$.

\subsection{Expansion by wavelets}

$F^j(x)$ contains less information than $F(x)$, because 
information on scales $\geq j+1$ (i.e. smaller scales) has been smoothed 
out. It would be nice not to lose any information during the smoothing 
process. This can be
accomplished if the differences,  $F^{j}(x)-F^{j-1}(x)$, between the 
smoothed distributions on successive scales are retained. In other words, 
only if we are able to retain all these {\it differences}, this scheme 
will not lose any information.

To calculate the differences, we define the difference function, or 
wavelet, as
%eq9
\begin{equation}
\psi^{H}(\eta) = \left (\frac{2^j}{L}\right)^{1/2}
  \left\{ \begin{array}{ll} 1 & \mbox{for $0 \leq
\eta \leq 1/2$} \\ -1 & \mbox{for $1/2 \leq \eta \leq 1$} \\ 0 &
\mbox{otherwise.}
\end{array} \right.
\end{equation}
This is the basic Haar wavelet. One can then construct a set of wavelets 
$\psi_{j,l}^{H}(x)$ by dilating and translating eq.(9) as
%eq10
\begin{equation}
\psi_{j,l}^{H}(x)  = \psi^{H}(2^{j} x/L - l).
\end{equation}
The Haar wavelets are orthonormal with respect to {\it both} indices 
$j$ and $l$, i.e.
%eq11
\begin{equation}
\int_0^L \psi_{j',l'}^{H}(x)\psi_{j,l}^{H}(x)dx =
 \delta^K_{j',j} \delta^K_{l',l} .
\end{equation}
For a given $j$, $\psi_{j,l}^{H}(x)$ is also orthogonal to the
scaling functions $\phi_{j',l}^{H}(x)$ for $j'\leq j$, i.e.
%eq12
\begin{equation}
\int_0^L \phi_{j',l'}^{H}(x)\psi_{j,l}^{H}(x)dx = 0, \hspace{2cm}
{\rm for \ \ \ } j' \leq j.
\end{equation}

 From eqs.(4) and (9), we have
%eq13
\begin{equation}
\begin{array}{ll}
\phi_{j,2l}^{H}(x) &
    = \phi_{j-1,l}^{H}(x) + \psi_{j-1,l}^{H}(x),\\
      &   \\
\phi_{j, 2l+1}^{H}(x) &
   = \phi_{j-1,l}^{H}(x) - \psi_{j-1,l}^{H}(x).\\
\end{array}
\end{equation}
Thus, the difference $F^{j}(x)-F^{j-1}(x)$ is given by
%eq14
\begin{equation}
F^{j}(x)-F^{j-1}(x)=
 \sum_{l=0}^{2^{j-1}-1} \tilde{\epsilon}^F_{j-1,l} \psi_{j-1,l}^{H}(x),
\end{equation}
where $\tilde{\epsilon}^F_{j-1,l}$ are called the wavelet function
coefficients (WFC) given by
%eq15
\begin{equation}
\tilde{\epsilon}^F_{j,l}=\int F(x)\psi_{j,l}^{H}(x)dx .
\end{equation}

Using the relation (14) repeatedly, we have
%eq16
\begin{equation}
F^{j}(x) = F^{0}(x)+
     \sum_{j'=0}^{j-1} \sum_{l=0}^{2^{j'}-1} \tilde{\epsilon}^F_{j',l}
 \psi_{j',l}^{H}(x).
\end{equation}
This is an expansion of the flux $F^{j}(x)$ with respect
to the basis $\psi_{j,l}^{H}(x)$, and $F^{0}(x)$ is the mean
of $F(x)$ on the entire range $L$. Therefore, the flux $F(x)$ can be 
can be expressed as
%eq17
\begin{eqnarray}
\lefteqn{ F(x) = F^{J}(x) } \\ \nonumber
  & & =F^{j}(x) + \sum_{j'=j}^{J-1} \sum_{l=0}^{2^{j'}-1}
  \tilde{\epsilon}^F_{j',l} \psi^H_{j',l}(x) \\ \nonumber
 & & = F^{0} + \sum_{j'=0}^{J-1} \sum_{l=0}^{2^{j'}-1}
  \tilde{\epsilon}^F_{j',l} \psi^H_{j',l}(x).
\end{eqnarray}

The Haar wavelet provides a clear picture of the DWT decomposition, 
and it is also easy for numerical work. However, the Haar wavelet is 
discontinuous in real space, and therefore, it is not well behaved 
in scale space. 
For our work, the most important properties of the basis for the 
scale-space decomposition are 1.) orthogonality, 2.) completeness, and 
3.) locality in both scale and physical spaces. 
Therefore, all wavelets with compactly supported basis will produce
similar results. Among the compactly supported orthogonal bases, 
the Daubechies 4 (D4) is easy for numerical calculation. We will use 
wavelet D4. 

For the D4, the basic 
orthonormal eqs.(5), (11) and (12) still hold after replacing 
$\phi^H_{jl}$ and $\psi^H_{jl}$ with D4 scaling function $\phi_{jl}$ 
and D4 wavelet $\psi_{jl}$. The 
expansion eqs.(17) are also valid using  $\phi_{jl}$ and $\psi_{jl}$. 

\subsection{The DWT power spectrum of flux without normalization}

 Form eq.(17), the Parseval theorem of the DWT gives
%eq18
\begin{equation}
\int[F(x)-F^{0}]^2dx = \sum_{j=0}^{J-1}\sum_{l=0}^{2^j-1}
    |\tilde{\epsilon}^F_{j,l}|^2.
\end{equation}
Therefore, the power of the mode $(j,l)$ is 
$|\tilde{\epsilon}^F_{j,l}|^2$. Thus, the power spectrum of the flux $F(x)$
is given by (Pando \& Fang 1998; Fang \& Feng, 2000; Yang et al. 2001)
%eq19
\begin{equation}
P^F_j = \langle |\tilde{\epsilon}^F_{j,l}|^2 \rangle
\end{equation}
The ensemble average of the power 
$\langle |\tilde{\epsilon}^F_{j,l}|^2 \rangle$ should be $l$-independent
for a homogeneous field. If the ``fair sample hypothesis'' is 
true (Peebles 1980), the 
ensemble average can be replaced by a spatial average, we then have
%eq20
\begin{equation}
P^F_j = \frac{1}{2^j}\sum_{l=0}^{2^j -1}  |\tilde{\epsilon}^F_{j,l}|^2.
\end{equation}
This is the DWT power spectrum of the flux $F(\lambda)$. It is a 
band-averaged Fourier power spectrum as
%eq21
\begin{equation}
P_j = \frac{1}{2^j} \sum_{n = -\infty}^{\infty}
 |\hat{\psi}(n/2^j)|^2 P(n),
\end{equation}
where $P(n)$ is the Fourier power spectrum with wavenumber $k=2\pi n/L$,
and $\hat{\psi}(n)$ is the Fourier transform of the basic
wavelet $\psi(x)$. This relation has been confirmed with 1-D 
(Ly$\alpha$) and 3-D (N-body) numerical samples (Feng \& Fang 
2000; Yang et al. 2001.)   

The DWT algorithm eq.(21) have been successfully employed for the power 
spectrum reconstruction (Feng \& Fang 2000; Pando, Feng \& Fang 2001). 
It can recover the initial linear power spectrum on small scales as well 
as large scale. For the reconstruction, the normalization is simple,
as the field is gaussian. For non-gaussian fields, the normalization
will no longer be trivial, as the non-gaussianity will affect the 
normalization differently for different algorithm.      

\section{Algorithm for the normalized DWT power spectrum}

\subsection{The DWT power spectrum from a Poisson sampling}

Using photon numbers $N(x) \propto F(x)$, eq.(2) can be rewritten as 
%eq22
\begin{equation}
\tilde{N}(x)=\overline{N(x)}[1+ \delta(x)]+ N_e(x).
\end{equation}
where $\overline{N(x)}=N_c(x)\overline{e^{-\tau(x)}}$ is the mean 
photon number at $x$ (wavelength or redshift). The term $N_e(x)$ 
is given by noise, which includes the background sky, dark current and 
the instrumental readout noise (see Appendix A). 

By definition, the mean photon number (or mean flux) is the 
background of the fluctuation $\delta(x)$. It corresponds to  
the selection function in the problem of galaxy distribution. 
Because $\overline{N(x)}$ or $\overline{F(x)}$ are not constant 
in the entire spatial range, the background can only be defined 
scale-by-scale. In other words, to measure the power of 
$\delta(x)$ on a given scale $j$, the background $\overline{N(x)}$ 
and $\overline{F(x)}$ are given by the mean photon number and mean 
flux at $x$ when all fluctuations on scales smaller than the given 
scale ($>j$) are absent, i.e.
%eq23
\begin{equation}
N(x)=\overline{N(l)}_j[1+ \delta(x)_j]+ N_e(x).
\end{equation}
and
%eq24
\begin{equation}
F(x)=\overline{F(l)}_j[1+\delta(x)_j]+ e(x).
\end{equation}
where $\delta(x)_j$ contains all information about the transmission
fluctuations on scales $\geq j$. $\overline{N(x)}_j$ and 
$\overline{F(l)}_j$ are the mean photon number and mean flux 
in the spatial range $lL/2^j < x < (l+1)L/2^j$, respectively, i.e.
the local mean photon number and mean flux when the mass field 
clustering on scales $ \geq j$ is absent. The error term in eq.(24) 
is $e({x})=N_e(x) F(x)/N(x)$ as $N(x)\propto F(x)$.

Due to the discreteness of photons, the observed photon number 
$N(x)$ should be considered as a sampling of the random field of
eqs.(22) or (23). Our purpose is to estimate the power spectrum of the 
random field $\delta(x)$, which describes the fluctuations 
of the transmission. If the sampling is Poissonian, 
eq.(22) yields
%eq25
\begin{eqnarray}
\lefteqn { \langle\delta({x})\delta({x'})\rangle =  
-1+ \left \langle \frac{N(x)N(x')}
  {\overline{N(x)}\ \overline{N(x')}}\right \rangle 
} \\ \nonumber
& & -  \delta^D(x-x')\left \langle\frac{1}{\overline{N(x)}} 
  \right \rangle -
\left \langle \frac{N_e({x})N_e({x'})}
{\overline{N(x)}\ \overline{N(x')}}\right \rangle.
\end{eqnarray}
The derivation of eq.(25) can be found in Appendix B which describes 
both Poisson and non-Poisson sampling. Considering $F(x)\propto N(x)$,
the second and fourth terms on the r.h.s. eq.(25) can be rewritten as
%eq26
\begin{eqnarray}
\lefteqn { \langle\delta({x})\delta({x'})\rangle =  
-1+ \left \langle \frac{F(x)F(x')}
  {\overline{F(x)}\ \overline{F(x')}}\right \rangle 
} \\ \nonumber
& & -  \delta^D(x-x')\left \langle\frac{1}{\overline{N(x)}} 
  \right \rangle -
\left \langle \frac{e({x})e({x'})}
{\overline{F(x)}\ \overline{F(x')}}\right \rangle.
\end{eqnarray}

 Projecting eq.(26) onto the DWT bases $\psi_{jl}(x)\psi_{jl}(x')$, 
the l.h.s. gives the DWT power spectrum of $\delta(x)$, i.e.
the {\it normalized} DWT power spectrum of the flux fluctuations,  
%eq27
\begin{equation}
P_j=\langle \tilde{\epsilon}^2_{jl} \rangle.
\end{equation}
The first term on the r.h.s. of eq.(26) disappears, because $\psi_{jl}(x)$ 
is admissible, i.e. $\int \psi_{jl}(x)dx=0$. Thus, eq.(26) yields
%eq28
\begin{equation}
P_j= \left \langle \left[
 \frac{\tilde{\epsilon}^F_{j,l}}{\overline{F(l)}_j}
    \right ]^2  \right \rangle - 
\left \langle \frac{1}{\overline{N(l)}_j}\right \rangle
-\left \langle \left[ 
\frac{\tilde{\epsilon}^E_{jl}}{\overline{F(l)}_j}\right ]^2
 \right \rangle.
\end{equation}
where $\tilde{\epsilon}^E_{jl}=\int e({x}) \psi_{j,l}(x)dx$.

Using again the ``fair sample hypothesis", we have 
%eq29
\begin{eqnarray}
\lefteqn{ P_j= \frac{1}{2^j}\sum_{l=0}^{2^j-1}
  \left [\frac{\tilde{\epsilon}^F_{j,l}}{\overline{F(l)}_j}
   \right ]^2} \\ \nonumber
& & - \frac{1}{2^j}\sum_{l=0}^{2^j-1}
   \frac{1}{\overline{N(l)}_j}
-\frac{1}{2^j}\sum_{l=0}^{2^j-1}
  \frac{1}{[\overline{F(l)}_j]^2}
    \int\sigma^2(x)\psi_{jl}^2(x)dx.
\end{eqnarray}
In calculating the error term, we used
%eq30
\begin{equation}
\langle e(\lambda)  \rangle=0, \hspace{1cm} 
\langle e(\lambda)e(\lambda') \rangle=\sigma^2(\lambda) 
  \delta_{\lambda, \lambda'}
\end{equation}
The standard derivation  $\sigma(\lambda)$ can be found from 
the data set 
given by observers. It is also reasonable to assume that the 
noise is uncorrelated with $\delta(x)$. 

Eq.(29) is the basic estimator of the normalized DWT power
spectrum of Ly$\alpha$ transmitted flux. The first term on the r.h.s. 
of eq.(29) is the power spectrum of the transmission. The second 
and third terms are the 
corrections due to shot noise and the noise $e(\lambda)$, respectively.

\subsection{The estimation of the mean flux}

Now we turn to the problem of estimating the mean flux $\overline{F(l)}_j$.
The definition of $\overline{F(l)}_j$ given by eq.(24) requires a 
decomposition of $F(x)$, in which the first term contains only information on 
scales larger than $L/2^j$, but nothing on scales less than or equal to  
$L/2^j$, and all the information of flux fluctuations on scales equal to or 
less than $L/2^j$ is contained in the second term. 

This decomposition is already given by second line of eq.(17), in which 
$F^{j}(x)$ is the flux at $x$ if all the fluctuations on scales less than or 
equal to  $L/2^j$ are absent. Thus, to measure the power on scale $j$, 
one can identify the mean flux $\overline{F(l)}_j$ with $F^{j}(x)$. 
Obviously, $F^{j}(x)$ is not constant, except within each small segment 
$lL/2^j  < x < (l+1)L/2^j$, as all fluctuations on scales $\geq j$ are 
smoothed out. Therefore, we have
%eq31
\begin{equation}
\overline{F(l)}_j= F^{j}(x) = 
 \sum_{l=0}^{2^j-1} \epsilon^F_{jl}\phi_{jl}(x),
\end{equation}
where, for a given $l$, $x$ is given by $lL/2^j  < x < (l+1)L/2^j$.  

Eq.(31) shows that the mean flux $\overline{F(l)}_j$ at position $l$ can 
be represented by the SFC $\epsilon^F_{jl}$ for scale $j$.
However, as mentioned in \S 2.1, the SFC $\epsilon^F_{jl}$ is 
proportional to, but not equal to the mean flux at position $l$. 
To find the proportionality constant, we use the so-called 
``partition of unity" of wavelets given (Daubechies 1992)
%eq32
\begin{equation}
\sum_{l=-\infty}^{\infty} \phi(\eta-l)=1,
\end{equation}
or
%eq33
\begin{equation}
\left (\frac{L}{2^j}\right)^{1/2} 
  \sum_{l=0}^{2^j-1}\phi_{jl}^P(x)=1,
\end{equation}
where $\phi_{jl}^P(x)$ is the so-called periodized scaling function 
(Fang \& Feng 2000). With $\phi_{jl}^P(x)$, eq.(7) can be rewritten as
%eq34
\begin{equation}
\epsilon^F_{j,l}=\int_{0}^{L} F(x) \phi^P_{j,l}(x)dx.
\end{equation}
Substituting eq.(3) into eq.(34), we have
%eq35
\begin{equation}
\epsilon^F_{j,l} =\sum_{i=1}^{n-1} \frac{1}{2}[F(\lambda_i)+F(\lambda_{i+1})]
   \int_{\lambda_i}^{\lambda_{i+1}}\phi^P_{j,l}(x)dx.
\end{equation}
Thus, with the ``partition of unity" eq.(32), eq.(35) yields
%eq36
\begin{equation}
\sum_{l=0}^{2^j-1} \left (\frac{L}{2^j}\right)^{1/2}
  \epsilon^F_{j,l} =\sum_{i=1}^{n-1}\frac{1}{2}[F(\lambda_i)+F(\lambda_{i+1})]
(\lambda_{i+1}-\lambda_{i}).
\end{equation}
This result is clear. The r.h.s. of eq.(36) is the integral flux of the
entire spectrum, while the l.h.s. is the integral flux of the smoothed
spectrum $F^{j}(x)$. That is, the smoothed spectrum $F^{j}(x)$ always 
has the same integral flux as the original spectrum. $F^{j}$ is only 
a reassignment of the original flux $F(x)$ from the 
distribution $F(\lambda_i)$ over grids $i=1..\ n$ to new distribution 
$\sqrt{L/2^j}\epsilon^F_{jl}$ over grids $l=0..\ 2^j-1$. 

The quantity $(L/2^j)^{1/2} \epsilon^F_{j,l}$ is the flux in the 
spatial range $lL/2^j$ to $(l+1)L/2^j$. Thus, the mean flux in this spatial 
range is
%eq37
\begin{equation}
\overline{F(l)}_j = \left(\frac{2^j}{L}\right)^{1/2}\epsilon^F_{jl}.
\end{equation}
Although the mean flux $\overline{F(l)}_j$ is different on different 
scales, the average of $\overline{F(l)}_j$ over $l$ is independent 
of $j$, i.e. 
%eq38
\begin{equation}
\frac{1}{2^j}\sum_{l=0}^{2^j-1}\overline{F(l)}_j 
 = \frac{1}{L}\sum_{i=1}^{n-1}\frac{1}{2}[F(\lambda_i)+F(\lambda_{i+1})]
(\lambda_{i+1}-\lambda_{i}).
\end{equation}
This allows comparison of powers on various scales. In other words, 
the power, $P_j$, on different scales $j$ is properly normalized.

 From the definition of the mean flux, $\overline{F(l)}_j$ can be 
written as
%eq39
\begin{equation}
\overline{F(l)}_j = F_c(l)_j [\overline{e^{-\tau(l)}}]_j.
\end{equation}
where $F_c(l)_j$ is the continuum at the cell $(j,l)$. 

If the continuum can be approximated as a constant, i.e. $F_c(l)_j=F_c$, 
eq.(38) gives
%eq40
\begin{equation}
\frac{1}{2^j}\sum_{l=0}^{2^j-1} \overline{e^{-\tau(l)}}_j =
  \frac{1}{F_c} \frac{1}{L}\sum_{i=1}^{n-1}
  \frac{1}{2}[F(\lambda_i)+F(\lambda_{i+1})]
(\lambda_{i+1}-\lambda_{i}).
\end{equation} 
Eq.(40) means that the average of the mean transmission 
$\overline{e^{-\tau(l)}}_j$ over positions $l$ is independent of $j$, 
i.e. it is a constant. However, even in this case, it doesn't mean
that the power spectrum can be normalized by a constant
mean transmission (\S 4).

\subsection{The estimator of normalized power spectrum of 
   transmission}

Substituting the mean flux eq.(37) into eq.(29), we have the 
transmission power spectrum estimator as 
%eq41
\begin{equation}
P_j= \frac{L}{2^{2j}}\sum_{l=0}^{2^j-1}
 \left[ \frac{\tilde{\epsilon}^F_{j,l}}{\epsilon^F_{jl}}\right]^2
   - \frac{1}{2^j}
   \sum_{l=0}^{2^j-1}\frac{1}{N(l)_j}
-\frac{L}{2^{2j}}\sum_{l=0}^{2^j-1}
  \frac{\overline{\sigma^2(l)}}{[\epsilon^F_{jl}]^2},
\end{equation}
where 
%eq42
\begin{equation}
\overline{\sigma^2(l)} = \int\sigma^2(x)\psi_{jl}^2(x)dx.
\end{equation}
It is the mean of $\sigma^2(l)$ at the  position $l$. 

In the algorithm of eq.(41), the fluctuation amplitudes 
$\tilde{\epsilon}^F_{j,l}$ (WFCs) at position $l$ are normalized by the
mean flux $\epsilon^F_{jl}$ (SFCs) at the same position $l$. 
That is, the normalization coefficients (SFCs) and the fluctuation amplitudes
(WFCs) are obtained from the same DWT decomposition eq.(17). We call it the 
self-normalized power spectrum estimator.  

Let consider the case of a constant $F_c$. Eq.(29) and 
(39) give
%eq43
\begin{equation}
P_j= \left \langle \left[
 \frac{\tilde{\epsilon}^F_{j,l}}{F_c\overline{e^{-\tau(l)}}_j}
    \right ]^2  \right \rangle - 
\left \langle \frac{1}{N(l)_j}\right \rangle
-\left \langle \left[ 
\frac{\tilde{\epsilon}^E_{jl}}{F_c\overline{e^{-\tau(l)}}_j}\right ]^2
 \right \rangle. 
\end{equation}
A constant $F_c$ actually doesn't affect the power $P_j$, because 
the power is given by the ratio $\tilde{\epsilon}^F_{j,l}/\epsilon^F_{jl}$ 
[eq.(40)], and both the $\tilde{\epsilon}^F_{j,l}$ and 
$\epsilon^F_{jl}$ are proportional to $F_c$ [eqs.(7) and (34)].

If there is no correlation between $\tilde{\epsilon}^F_{j,l}$ 
and $\overline{e^{-\tau(l)}}_j$, eq.(43) gives
%eq44
\begin{equation}
P_j= \left \langle 
 \frac{1}{|\overline{e^{-\tau(l)}}_j|^2} \right \rangle
  \left \langle  \left [  
   \frac{\tilde{\epsilon}^F_{j,l}}{F_c}\right ]^2  \right \rangle - 
\left \langle \frac{1}{N(l)_j}\right \rangle
-\left \langle \left[ 
\frac{\tilde{\epsilon}^E_{jl}}{F_c\overline{e^{-\tau(l)}}_j}\right ]^2
 \right \rangle. 
\end{equation}
 From eq.(40), the factor 
$\langle 1/|\overline{e^{-\tau(l)}}_j|^2 \rangle$ is approximately
independent of $j$. That is, the normalization of power spectrum (44) 
is given by a constant mean transmission. If the WFCs 
$\tilde{\epsilon}^F_{j,l}$ are correlated
with $\overline{e^{-\tau(l)}}_j$ or the SFCs $\epsilon^F_{j,l}$, the 
power spectrum cannot be normalized by a constant mean transmission
$\overline{e^{-\tau}}$, even when the continuum is constant.

\section{Test of the self-normalization algorithm}

In this section, we show that the power spectrum given by eq.(41) is
properly normalized. It doesn't need a continuum-fitting, or a 
pre-calculation of the mean transmission. 

\subsection{Simulation samples of the lognormal model}

Our purpose is only to demonstrate the normalization, but not model 
discrimination or cosmological parameter determination. We can use 
the semi-analytical model, the so-called lognormal model developed
by Bi \& Davidsen (1997). The lognormal distribution is useful 
for testing algorithm on intermittent fields, as lognormal field is 
typically intermittent.

Briefly, this simulation consists of three steps. First, we generate 
a realization of the primordial (linear) baryonic mass 
distribution of size 200 h$^{-1}$ Mpc in comoving space, centered at 
a typical redshift, say $z=2.4$ in $\Omega =1$ cold dark matter 
universe. The COBE data is used to normalize the initial spectrum 
which has the soften parameter $\Gamma = 0.3$. The Hubble constant is 
taken to be 50 km s$^{-1}$ Mpc$^{-1}$. In the Fourier space, the 
distribution of baryonic matter differs from that of dark matter due to 
smoothing on the Jeans length. Second, the non-linear 
baryonic density is calculated 
using the lognormal transformation from the linear density; and the  
neutral baryonic density is calculated assuming an UV ionization flux of 
$J_0 = 1.0\times 10^{-22}$ erg cm$^{-2}$ s$^{-1}$ Hz$^{-1}$ sr$^{-1}$. 
The temperature of the gas is assumed to have a mean of 
$2.0\times 10^4$ K following an adiabatic equation of state with the 
polytropic index $4/3$. Because of the UV radiation, a minimal temperature 
should be introduced. We take it to be $1.12\times 10^4$ K. Lastly, the 
absorption of the Ly$\alpha$ photons by hydrogen is convolved by 
a Voigt profile, and the whole optical depth is calculated by 
summing over all the pixels. Each pixel has the size of 0.0156$\AA$ and 
the total number of pixels is 32768.

To fit with the observed data of medium resolution, we  
convolve the theoretical spectrum by an instrumental 
point-spread-function (PSF) of the typical resolution 41 km s$^{-1}$. 
Alternatively, one can simulate a medium resolution spectrum by using 
a higher gas temperature but without the instrument PSF. 
We will use the temperature $1.0\times 10^5 K$ to mimic the instrumental
effect. With this simulation, we have the transmission 
$e^{-\tau(\lambda)}$, which is shown in the top panel of Fig. 1.

We then assume the CCD counts $N_c=2000$ per wavelength if the 
fluctuations are absent. Three continua $F_c$ are considered: 1) flat 
continuum, $F_c=N_c$, 2) $F_c=N_c(\lambda / \lambda _0)^2$, where 
$\lambda _0$ is $4133$ \AA (redshift 2.4), 3) 
$F_c=N_c[1+ d \sin [2\pi(\lambda-\lambda_{min})/(\lambda_{max}-
\lambda_{min})]$, and $\lambda_{min}= 3890.49$ \AA, 
$\lambda_{max}= 4399.53$ \AA. The amplitudes $d$ is taken to be 0.1, 0.2,
0.3, 0.4 and 0.5. The observed photons are a Poisson sampling of a 
random field with mean $F_ce^{-\tau}$. For each given $F_c$, hundred 
realizations of the Poison sampling are produced. Finally, gaussian 
noise with zero mean counts and standard derivation 50 (photon number),
which is independent from the Poisson sampling, is added to each 
pixel. Some results are shown in Fig. 1.

\subsection{Continuum-independent power spectrum}

Using the estimator in eq.(41), we calculate the DWT power spectra 
from the simulated flux with various continua. The results are plotted 
in Fig. 2. It shows that the DWT power spectra for different continua 
are exactly the same on scales $j\geq 3$ (or length scale 
$2^{15-j}\times 2$ h$^{-1}$ kpc). The dispersion of the power spectra over 
100 realizations is very small. That is, the power spectrum given 
by estimator (41) is continuum-independent. That it, the self-normalization
algorithm can produce the power spectrum correctly normalized by
the continuum, but without a continuum fitting.

\subsection{Normalization of mean transmission}

To test the mean transmission normalization of estimator eq.(41),
we calculate the so-called unnormalized DWT power spectrum of the 
transmission, i.e. the power spectrum of continuum normalized flux 
%eq45
\begin{equation}
e^{-\tau(\lambda)} = \frac{F(\lambda)}{F_c(\lambda)}.
\end{equation}
Similar to eq.(20), the unnormalized DWT power spectrum is given by
%eq46
\begin{equation}
P^t_j=\frac{1}{2^j}\sum_{l=0}^{2^j-1} |\tilde{\epsilon}^t_{jl}|^2
\end{equation}
where the WFCs are calculated by 
%eq47
\begin{equation}
\tilde{\epsilon}^t_{jl}=\int \frac{F(\lambda)}{F_c(\lambda)} \psi_{jl}(x)dx.
\end{equation}
The result of $P^t_j$ for the lognormal sample is plotted in Fig. 2. 

By definition eq.(1), we have
%eq48 
\begin{equation}
\delta(\lambda)=
 \frac {e^{-\tau(\lambda)}}{\overline {e^{-\tau(\lambda)}}}-1
\end{equation}
The second term on the r.h.s. of eq.(48) does not contribute to the DWT 
power spectrum, as $\psi_{jl}(x)$ is admissible. Therefore, the 
power spectrum $P_j^t$ should be transformed to $P_j$ by dividing a 
normalization factor $[\overline {e^{-\tau}}]^2$, or
%eq49
\begin{equation}
\overline{e^{-\tau(\lambda)}}_j=[P^t_j/P_j]^{1/2}.
\end{equation}
One can test the normalization of mean transmission by comparing
the value of $\overline{e^{-\tau(\lambda)}}_j$ given by eq.(49)] 
with the value $\overline{e^{-\tau}}$ used in the simulation. 
For the lognormal simulation samples, the result is shown in Fig. 3, 
in which three samples with different $\overline{e^{-\tau}}$ are analyzed.
It shows that $\overline{e^{-\tau(\lambda)}}_j$ on small $j$ (large 
scales) is exactly the same as the simulation used $\overline{e^{-\tau}}$.
That is, the estimator eq.(41) is able to produce the power spectrum, which 
is already properly normalized by the mean transmission.
Thus, the estimator eq.(41) doesn't need one to pre-calculate the mean 
transmission or filling factor for the normalization. 

However, on small scales, $\overline{e^{-\tau(\lambda)}}_j$ doesn't 
equal to $\overline{e^{-\tau}}$. Therefore, on small scale, the 
estimator eq.(41) is not the same as the traditional normalization by 
a mean transmission. This difference will be analyzed in next section.  

\section{Normalization of power spectrum of a non-linear field}

\subsection{Effective mean transmission}
 
We call $\overline{e^{-\tau(\lambda)}}_j$ the effective mean transmission.
Fig. 3 shows that the effective mean transmission  
$\overline{e^{-\tau(\lambda)}}_j$ is scale-dependent. 
It means that the power spectrum given by eq.(41) actually is normalized
by a scale-dependent factor $\overline{e^{-\tau(\lambda)}}_j$, not by
a mean transmission. Generally, $\overline{e^{-\tau(\lambda)}}_j$ is lower
than the mean transmission $\overline{e^{-\tau}}$ on small scales. 

The scale dependence of the effective mean transmission is due to
the anti-correlation between the background $\overline{e^{-\tau(\lambda})}$
and the fluctuation $\delta(\lambda)$. For a lognormal sample, this 
anti-correlation is shown in Fig. 4. The mean value of the WFCs
$\tilde{\epsilon}^F_{jl}$ (fluctuation) is larger for smaller SFCs 
$\epsilon^F_{jl}$ (background), and vice versa. On the other hand, 
the power is dominated by large WFCs, corresponding to smaller SFCs. 
Thus the normalization by an over-all averaged SFCs, or mean 
transmission $\overline{e^{-\tau}}$ will produce lower power  
than the power spectrum given by eq.(41). This anti-correlation 
comes from the spiky structures of an intermittent field. It is 
mostly significant on scale $j= 8$ and 9 (100 - 250 h$^{-1}$ kpc). 
On scale $j=10$ (or 50 h$^{-1}$ kpc), the anti-correlation is weak 
due to noise.

On the other hand, the field on large scales basically is gaussian, i.e.
no correlation between the WFCs $\tilde{\epsilon}^F_{jl}$ and the SFCs 
$\epsilon^F_{jl}$. In this case eq.(44) shows that the normalization can 
be done by a constant mean transmission $\overline{e^{-\tau(\lambda)}}$. 
This is why the mean transmission can be recovered by the ratio eq.(49) 
on large scales [Fig. 3].
             
\subsection{Test by pseudo-hydro simulation samples}

In the lognormal model, the non-linear feature of the field is induced  by 
the lognormal transform. To further test the normalization of effective 
mean transmission, we use simulation samples of the Ly$\alpha$ forests 
produced with the so called "pseudo-hydro" technique (Croft et al. 1998). 
In this approach, the non-linear density and velocity field of 
underlying dark matter 
were obtained by evolving a particle mesh (PM) simulation for a 
specified cosmological model. The gas density and temperature were 
then computed using the simple scaling relation inferred from full 
hydrodynamic simulations (Hui \& Gnedin 1997).

The parameters of the simulation are taken to match the Keck HIRES 
spectra described in Croft et al (2000). First, the PM simulations 
were performed by evolving $128^3$ dark matter particles in a 
periodic box with a $128^3$ grid. For this study, the cosmological 
model is taken to be the low density flat model (LCDM), which was 
specified by the density parameter $\Omega_0=0.3$, the cosmological 
constant $\Omega_{\Lambda}=0.7$, the Hubble constant $h=0.7$ and 
the baryon density parameter $\Omega_b=0.0125h^{-2}$. The physical 
size of box was determined correspondingly by 512 pixels around 
the median redshift $z_{med}$ of the sample, and a total of 100 time 
steps were integrated from the initial redshift $z=25$ to $z_{med}$.

We then select random lines-of-sight through the simulation box along
which to interpolate one-dimensional density and velocity field
using the Daubechies-4 scaling functions with j=7. Using
one-dimensional density field, we assign temperature to each
pixel using the polytropic equation of state $T=T_0\rho^{\alpha}$, 
where $T_0$ and $\alpha$ depend on the spectral shape of the UV 
background and on the history of reionization. We adopt typical 
values $T_0 \approx 10^4$K and $\alpha=0.6$. The neutral hydrogen 
fraction in each pixel is computed by adopting the cosmic abundance 
of hydrogen and assuming photoionization equilibrium. The optical depth
$\tau$ at a given pixel is then obtained by integrating in real
space by including the effect of peculiar velocity field and
convolving with Voigt thermal broadening profile. To match the parameters
of observed spectra, $\tau$ was computed onto a $2^9=512$ grid and 
smoothed by a gaussian window to match with the spectral resolution 
of observations. The absorption transmitted spectra $F=\exp(-\tau)$ is 
normalized such that the mean transmission in the spectra matches with 
observations. We take 29 different values of the mean transmission, 
at different redshifts. 

We calculated the effective mean transmission 
$\overline{e^{-\tau(\lambda)}}_j$ [eq.(49)] for the pseudo-hydro
simulation samples. As an example, we show a result of 
$\overline{e^{-\tau(\lambda)}}_j$ vs. $j$ for a simulated sample
in Fig. 5. It is similar to Fig. 3, i.e. the effective mean 
transmission on small scales is lower than the mean transmission.
Fig. 6 gives 29 $\overline{e^{-\tau(\lambda)}}_j$ on scale 
$k\simeq 0.02$ h Mpc$^{-1}$, and comparing them with the 29
$\overline{e^{-\tau}}$ used for the simulation. It shows that 
all the effective mean transmission on this small scale is lower 
than the corresponding mean transmission used for simulation. This 
result is the same as lognormal model. 

Therefore, the self-normalized power spectrum estimator eq.(41) 
is sensitive to both the mean transmission (on large scales) and 
the anti-correlation between the fluctuation and background on small 
scales. The power spectrum given by traditional normalization
is only sensitive to the mean transmission, but not the 
fluctuation-background anti-correlation on small scales.  
Therefore, the estimator eq.(41) would be more useful for 
the discrimination between models of Ly$\alpha$ forests. We will 
demonstrate this point in next subsection.  

\subsection{The power spectrum of a real sample} 

To demonstrate the estimator eq.(41),  we now analyze a real data set: 
the Ly$\alpha$ forest spectrum of q1700+642 given by Dobrzycki 
and Bechtold (1996). The data include the flux and 
continuum in the wavelength range from 3731.04 to 4611.26 \AA \ with 
resolution $\sim 0.25$ \AA. We use the 2$^{11}$ pixels from 3816.81 to 
4080.38 \AA. We remove regions in the spectra where the signal to 
noise ratio is less than 2.0. 

Using the estimator in eq.(41), we calculated the normalized DWT power 
spectrum $P_j$. From the data of flux and continuum, one can also 
calculate the unnormalized power spectrum $P^t_j$ by eq.(46). Figure 7 
shows the power spectra $P_j$ and $P^t_j$. It also shows the 
traditionally normalized power spectrum, i.e. $P^t_j/\overline{e^{-\tau}}$.

Similar to Fig. 2, Fig. 7 shows that the shapes of the two power 
spectra $P_j$ and $P^t_j$ of q1700+642 are completely the same on 
large scales ($j\leq 6$, or $\geq$ 1.1 h$^{-1}$ Mpc). In this scale
range, the power spectrum $P_j$ is perfectly the same as the traditionally
normalized power spectrum (see the coincidence of the solid lines and 
dot-dashed line). It means that the estimator eq.(41)  gives correct 
normalization on large scales. The ratio of $P_j/P^t_j$ on large scales 
($j \leq 6$, or $\geq$ 1.1 h$^{-1}$ Mpc) is $\simeq 1.4$, and therefore 
the mean transmission is $\overline{e^{-\tau}}\simeq 1/\sqrt{1.4} = 0.845.$ 
This value is exactly the same as the $\overline{e^{-\tau}}$ given by 
direct measurement. 

On small scales $j>6$ (1.1 h$^{-1}$ Mpc), the anti-correlation between 
the background and fluctuations becomes significant. Fig. 8 shows that 
the WFC- SFC anti-correlation is stronger on scales $j=8$ and 9 
(100 - 300 h$^{-1}$ kpc). In this scale range, the self-normalized $P_j$ is
generally larger than that given by traditional normalization, i.e. with
a constant mean transmission. At $j=9$ ($\simeq$ 140  h$^{-1}$ kpc), 
the traditionally normalized power spectrum is lower than the self-normalized
$P_j$ by a factor of 5.6.

What does this difference indicate? We know that the WFC-SFC 
anti-correlation is due to the intermittency of the field. In terms of 
Fourier decomposition, an intermittent field means that the phases of the 
Fourier modes is highly correlated, but the one point distribution of the 
Fourier amplitude is still gaussian due to the central limit theorem.
Therefore, the WFC-SFC anti-correlation can be eliminated by a 
phase-randomization, which is produced by taking the inverse transform of 
the Fourier coefficients of the original data after randomizing their 
phases uniformly over $[0,2\pi]$ without changing their amplitudes.
The phase randomized sample gets rid of the intermittent behavior 
possessed by the field, but the unnormalized power spectrum $P_j^t$ and 
the traditional normalized power spectrum of the phase randomized sample 
are exactly the same as the original one (Jamkhedkar, Zhan \& Fang 2000;
Zhan, Jamkhedkar \& Fang 2001). In 
Fig. 7, the long dashed line is the unnormalized power 
spectrum of the original data and its phase randomized counterpart, 
and the dot dashed line is the traditionally normalized power spectrum  
of the original data and its phase randomized counterpart.

Therefore, neither unnormalized nor traditionally normalized 
power spectrum can distinguish between the highly intermittent field 
and its phase-randomized counterpart.  
 
On the other hand, the estimator eq.(41) can detect the difference 
between the two fields. We calculate the power spectrum of the phase 
randomized sample by the self-normalized estimator eq.(41). The result 
is shown in Fig. 7. It shows that the self-normalized power spectrum 
of the original data is very different from its phase-randomized 
counterpart.  
 
Therefore, one can conclude that the self-normalized power spectrum 
estimator is sensitive to both the clustering behaviors of the 
field on large scales (mean transmission) and on small scales 
(intermittency, or fluctuation-background anti-correlation). But the
traditional normalized power spectrum is insensitive to the phase 
correlation of the Fourier modes.   

\section{Conclusion}

The power spectrum of Ly$\alpha$ forests is a direct indicator of the 
matter distribution at high redshift. This paper addresses the issue of 
how continuum fitting and the mean transmission affect the estimated 
power spectrum of QSO's Ly$\alpha$ forests. We propose a straightforward 
method for calculating the power spectrum of observed Ly$\alpha$ forests. 
This method is based on the DWT decomposition of the transmission flux. 
It gives a consistent calculation for the decomposition of flux and the 
normalization of power spectrum. 

With numerical simulation samples, we showed that the power spectrum 
obtained by this estimator is independent of the continuum. The 
non-linear power spectrum of the transmission can be reliably recovered 
from the observed flux regardless of the continuum, i.e. the algorithm
can automatically take care of the normalization by the  continuum without
a continuum fitting.

With numerical simulation samples, we also show that the power 
spectrum estimator can automatically consider the normalization
of the mean transmission, i.e. the algorithm doesn't need
a pre-calculated mean transmission to do the normalization.

For a gaussian field, the power spectrum given by the proposed 
estimator principally is the same as the power spectrum given by 
traditional normalization. In this case, an advantage of the 
proposed estimator is that it is free from fitting parameters. 

On scales with significant non-linear clustering, like 
intermittency or phase correlation, the self-normalized power spectrum 
is essentially different from the power spectrum normalized by 
traditional method. The latter is not sensitive to the phase
correlation, while the former is. Therefore, as an estimator of 
power spectrum of non-linear field traced by Ly$\alpha$ forests, the 
self-normalization algorithm is useful of the discrimination among 
models. 

\acknowledgments

We thank Drs. L.L. Feng and W.L.Lee for their help. PJ would 
also like to thank Jennifer Scott for useful discussions.

\appendix

\section{Photon counts and flux} 

The observed photon counts, $N(\lambda)$, at a pixel corresponding
to wavelength $\lambda$ is given by
%eqA1
\begin{equation}
N_{ob}(\lambda)=C(\lambda)N_c(\lambda) e^{-\tau(\lambda)}+ E(\lambda)
\end{equation}
where $\tau(\lambda)$ is the optical depth, and $N_c$ is the
continuum. $C(\lambda)$ describes the $\lambda$-dependence of
CCD's efficiency. $E(\lambda)$ is noise, whose mean and 
covariance are
%A2 
\begin{equation}
\langle E(\lambda)  \rangle=\bar{E}(\lambda), \hspace{1cm}
\langle E(\lambda)E(\lambda') \rangle=\sigma^2_{E}
\delta^K_{\lambda, \lambda'},
\end{equation}
where $\delta^K_{\lambda, \lambda}$ is Kronecker delta function.
Eq.(A2) means that the random variable $E(\lambda)$ is independent 
for each wavelength $\lambda$. 

The observed photon counts are reduced as  
%A3
\begin{equation}
N(\lambda)= \frac{N_{ob}(\lambda)- \bar{E}(\lambda)}{C(\lambda)}.
\end{equation}
We have then
%A4
\begin{equation}       
N(\lambda)=N_c(\lambda)e^{-\tau(\lambda)}- \frac{1}{C(\lambda)}
[\bar{E}(\lambda)-E(\lambda)].
\end{equation}
If we define a new variable for error as
%A5
\begin{equation}
N_e(\lambda) = \frac{1}{C(\lambda)}[\bar{E}(\lambda)-E(\lambda)],
\end{equation}
we have
%A6 
\begin{equation}
N(\lambda)=N_c(\lambda)e^{-\tau(\lambda)}-N_ e(\lambda),
\end{equation}
where
%A6
\begin{equation}
\langle N_e(\lambda)  \rangle=0, \hspace{1cm} 
\langle N_e(\lambda)N_e(\lambda') \rangle=\sigma'^2(\lambda) 
  \delta^K_{\lambda, \lambda'}
\end{equation}
The fluctuation of the transmission is defined by 
%A7
\begin{equation}
\delta(\lambda)=\frac{e^{-\tau(\lambda)} - \overline{e^{-\tau(\lambda)}}}
                  {\overline{e^{-\tau(\lambda)}}}.
\end{equation}
Thus, eq.(A6) yields eq.(21). 

\section{Poisson sampling and modified Poisson sampling}

Consider the reduced photon count $N(x)$ as a sampling of random field 
%eqB1
\begin{equation}
\tilde{N}(x)=\bar{N}(x)[1+\delta(x)],
\end{equation}
where $\bar{N}(x)=N_c\overline{e^{-\tau(\lambda)}}$. 
For the Poisson sampling, the characteristic function of $\tilde{N}(x)$ is
%eqB2
\begin{equation}
Z[e^{i\int \tilde{N}(x)u(x)dx}]= \exp\left \{ 
   \int dx\tilde{N}(x)[e^{iu(x)}-1] \right \}.
\end{equation}
Thus, the correlation functions of $N(x)$ are given by
%eqB3
\begin{equation}
\langle N(x_1)...N(x_n)\rangle_P =\frac{1}{i^n} \left [
\frac {\delta^n Z}{\delta u(x_1)... \delta u(x_n)} \right ]_{u=0},
\end{equation}
where $\langle ...\rangle_P$ is the average for the Poisson
sampling. We have then
%eqB4
\begin{equation}
\langle N\rangle_P = \tilde{F}(x),
\end{equation}
and
%eqB5
\begin{equation}
\langle N(x)N(x')\rangle_P = \tilde{N}(x)\tilde{N}(x') +
\delta^D(x-x')\tilde{N}(x).
\end{equation}
This equation yields
%B6
\begin{equation}
\langle \delta( x)\delta( x')\rangle= -1+ \left \langle
\frac{\langle N(x)N(x')\rangle_P}
  {\bar{N}(x)\bar{N}(x') } \right \rangle
- \delta^D(x-x')\frac{1}{\bar{N}(x)}.
\end{equation}
This gives eq.(24).

For a weighted Poisson sampling, the data at $x$ are given as a 
Poisson sampling of $\tilde{N}(x)$, but with a weight $g(x)$. 
In this case, the characteristic function eq.(B2) becomes
%B7
\begin{equation}
Z[e^{i\int\rho^g(x)u(x)dx}]= \exp\left \{ \int dx
\tilde{N}(x)[e^{ig(x)u(x)}-1] \right \}.
\end{equation}
We have then
%B8
\begin{equation}
\langle N\rangle_P = g(x)\tilde{N}(x),
\end{equation}
and
%B9
\begin{equation}
\langle N(x)N(x')\rangle_P = g(x)g(x')\tilde{N}(x)\tilde{N}(x') +
\delta^D(x-x')g^2(x)\tilde{N}(x).
\end{equation}
Eq.(B2) yields
%B10
\begin{equation}
\langle \delta({\bf x})\delta({\bf x'})\rangle= -1+ \left \langle
\frac{\langle N(x)N(x')\rangle_P}
  {g(x)\bar{N}(x)g(x)\bar{N}(x') } \right \rangle
- \delta^D(x-x')\frac{1}{\bar{N}(x)}.
\end{equation}

\newpage

\figcaption{A simulation sample of the QSO Ly$\alpha$ forests. The 
transmission $e^{-\tau}$ is shown in top panel. The lower four  
panels are the Poisson sampling of flux $F_c e^{-\tau}$
plus noise  with 1.) $F_c=2000$; 2.) $F_c=2000(\lambda / \lambda _0)^2$, 
$\lambda _0 =4133\AA$, 3.)
$F_c=2000[1+ 0.1 \sin [2\pi(\lambda-\lambda_{min})/(\lambda_{max}-
\lambda_{min})]$; and
4.) $F_c=2000[1+ 0.5 \sin [2\pi(\lambda-\lambda_{min})/(\lambda_{max}-
\lambda_{min})]$, respectively.}

\figcaption{Upper panel: The DWT power spectra $P_j$ of the simulated 
Ly$\alpha$ forest samples with continuum as 1.) $F_c=2000$ (solid line); 
2.) $F_c=2000(\lambda / \lambda _0)^2$ (dotted line),  where 
$\lambda _0 =4133$ \AA; and 3.) 
$F_c=2000[1+ 0.1 \sin [2\pi(\lambda-\lambda_{min})/(\lambda_{max}-
\lambda_{min})]$ (short-dotted line), respectively. 
Lower panel: The DWT power spectra $P_j$ of the simulated Ly$\alpha$ 
forest samples with continua given as  
$F_c=2000[1+ d \sin [2\pi(\lambda-\lambda_{min})/(\lambda_{max}-
\lambda_{min})]$ and $d=$ 0.1 (solid), 0.2 (dotted), 0.3(short dashed),
0.4 (long dashed) and 0.5 (dot-dashed line), respectively. 
The short-dash long-dashed line is the power spectrum of the 
transmission, $P_j^t$. 
The 1-sigma error bars are calculated from 100 samples. The scale $j$ 
corresponds to 2$^{15-j}\times 2$ h$^{-1}$ kpc.}

\figcaption{The effective mean transmission 
$\overline{e^{-\tau(\lambda)}}_j$ vs. $j$ (circle) 
[eq.(49)] for three lognormal simulation samples. The solid line is the mean 
transmission used for the simulation.
}

\figcaption{The histogram showing the number distribution of the SFCs. 
Their values can be read from the right hand $y$ axis. The 
squares show the average value of the |WFCs| in each SFC bin. Their 
values can be read from the left hand $y$ axis. The scale $j$ corresponds 
to 2$^{15-j}\times 2$ h$^{-1}$ kpc.}

\figcaption{The effective mean transmission 
$\overline{e^{-\tau(\lambda)}}_j$ vs. $j$ (circle) 
[eq.(49)] for a pseudo-hydro simulation samples. The solid line is the mean 
transmission used for the simulation.
}

\figcaption{The effective mean transmission 
$\overline{e^{-\tau(\lambda)}}_j=\sqrt{P^t_j/P_j}$ vs. the mean
transmission $\overline{e^{-\tau}}$ used for the 29 pseudo-hydro simulations.
The solid line is for $\overline{e^{-\tau}}.$
}

\figcaption{The transmission power spectrum $P_j$ (solid line) for the real 
sample q1700+642, calculated by estimator (41), and the power spectrum $P_j^t$ 
(long dashed line) calculated by eq.(46). The dot-dashed line represents 
the power spectrum normalized by traditional method. The dotted line is 
the power spectrum by eq.(46) for the phase randomized field. The error bar 
is given by the maximum and minimum of 20 realizations.  The 
scale $j$ corresponds to  2$^{11-j}\times 34$ h$^{-1}$ kpc.}

\figcaption{The WFC-SFC anti-correlation for real sample of the q1700+642 
Ly$\alpha$  forests. The histogram showing the number distribution of the 
SFCs. Their values can be read from the right hand $y$ axis. The 
squares show the average value of the WFCs in each SFC bin. Their 
values can be read from the left hand $y$ axis. The scale $j$ corresponds to 
2$^{11-j}\times 34$ h$^{-1}$ kpc.}

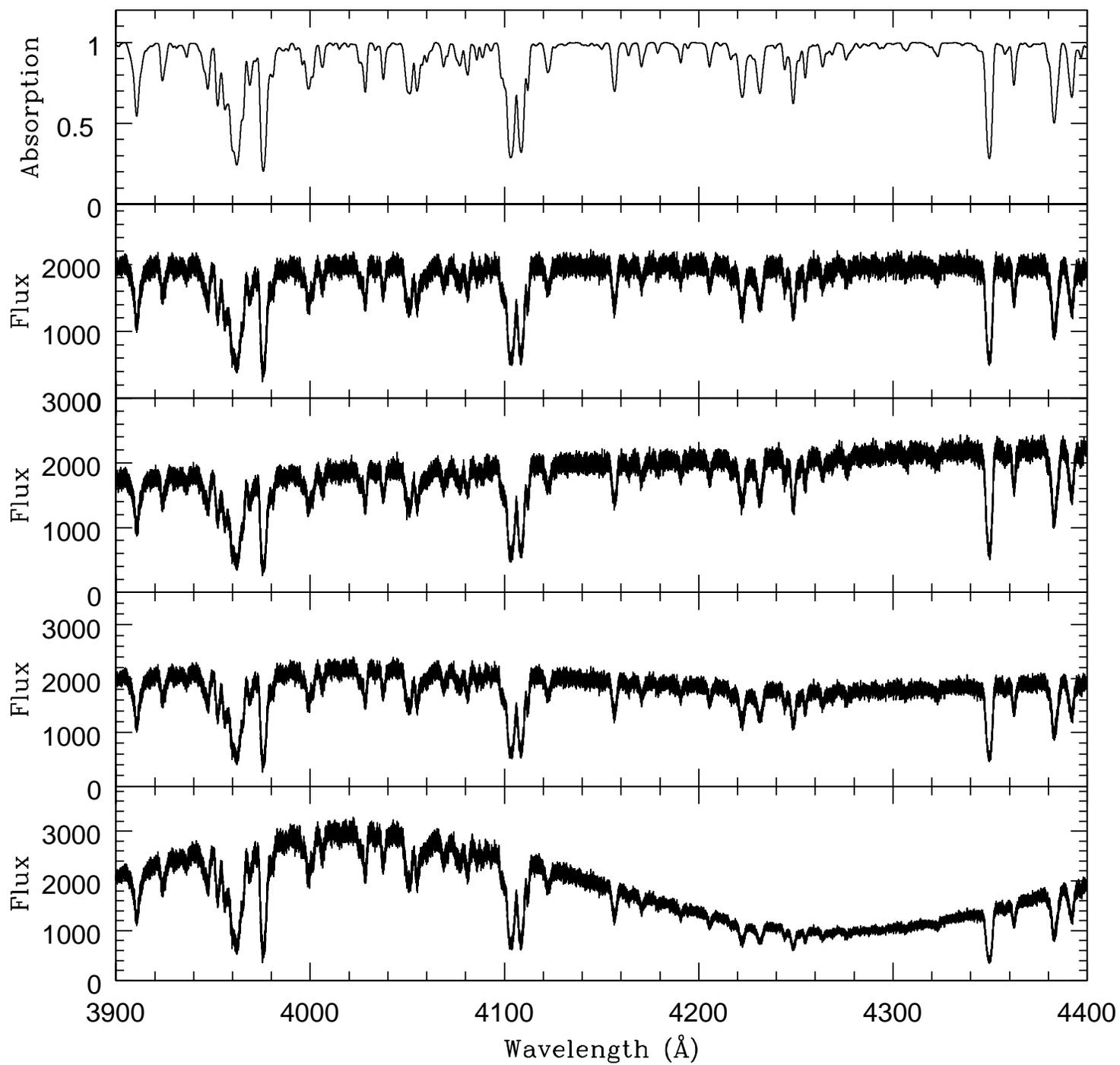
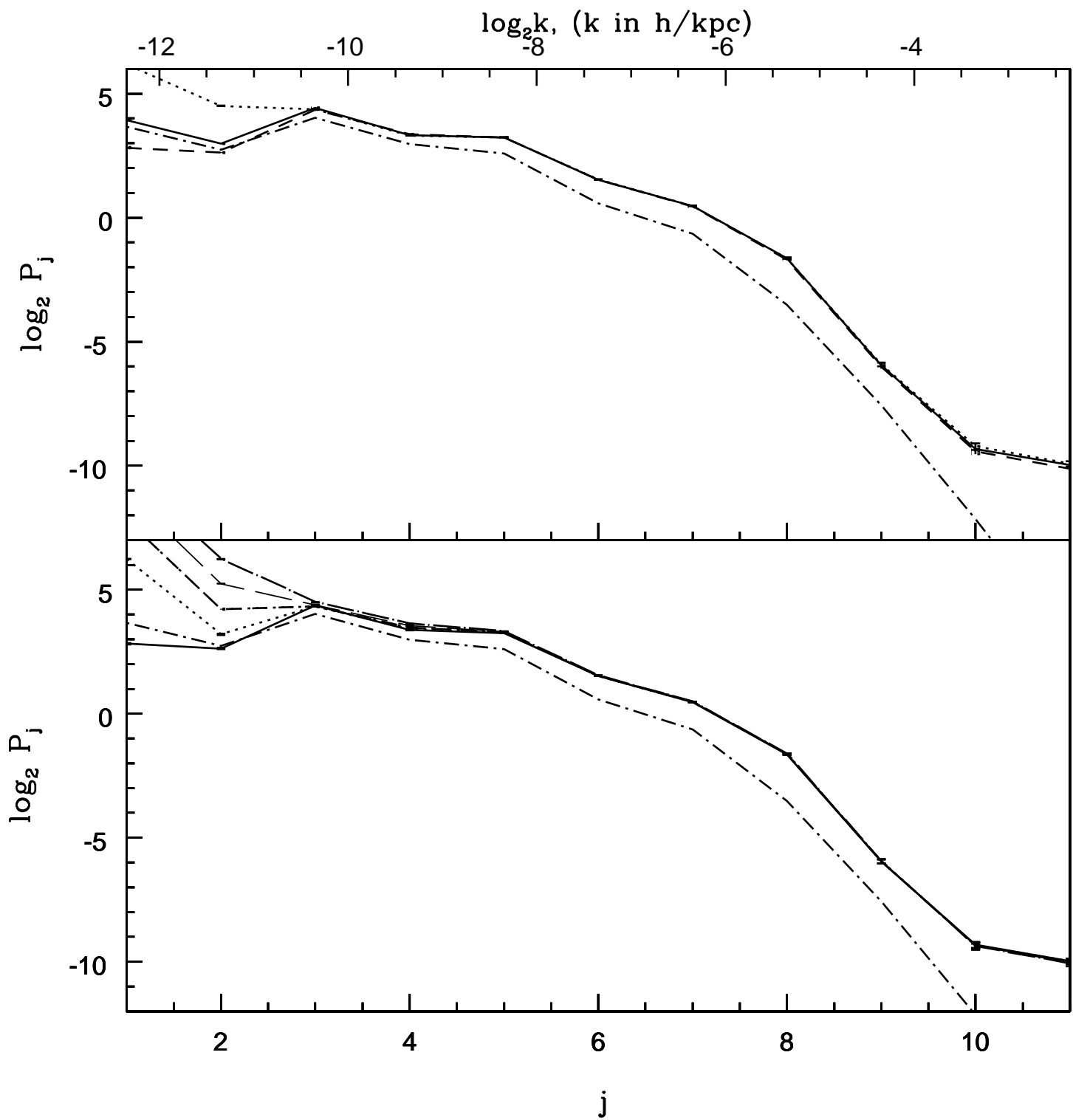
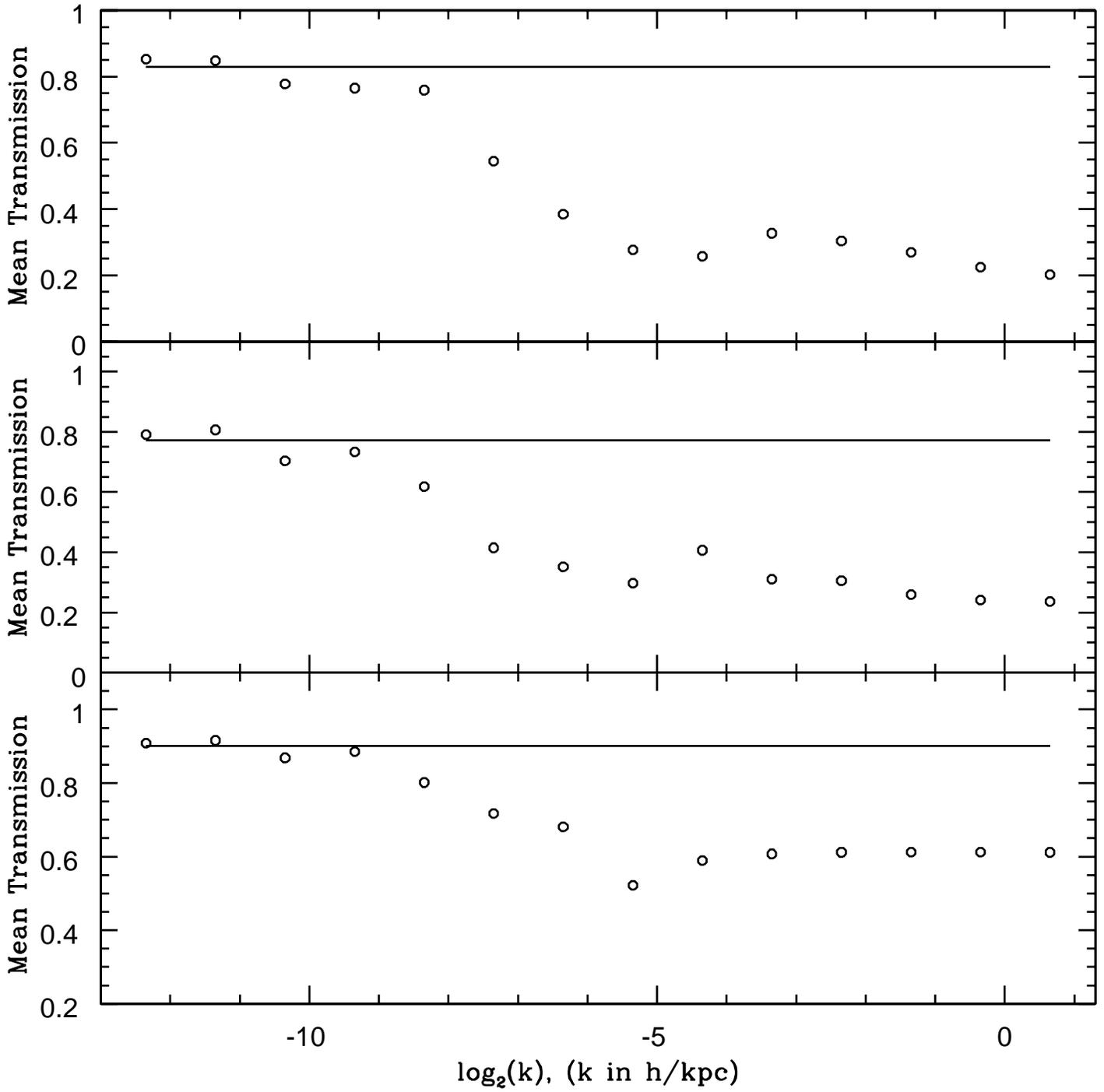
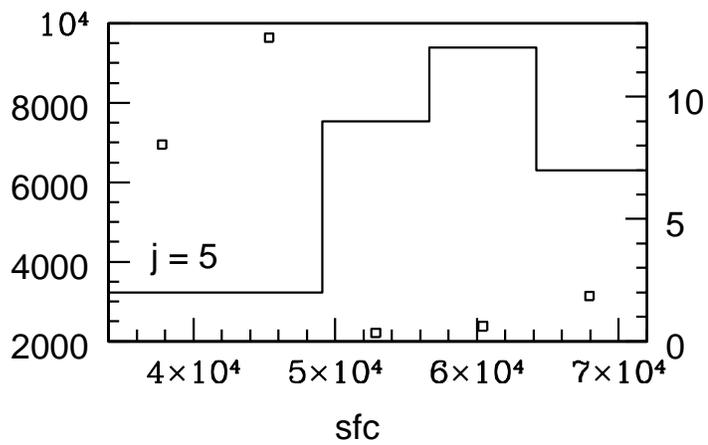
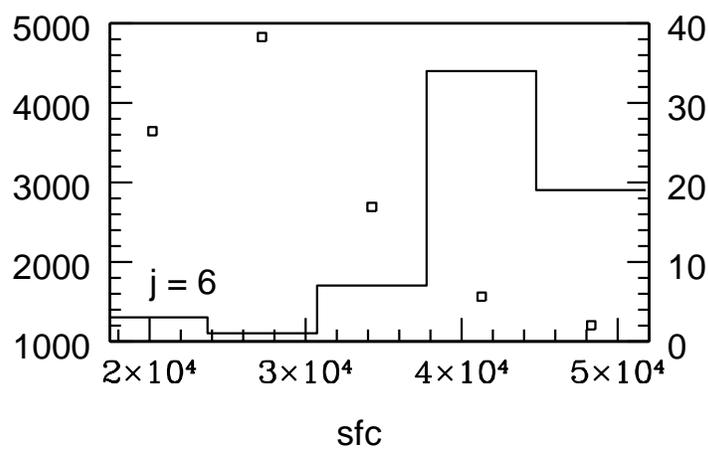
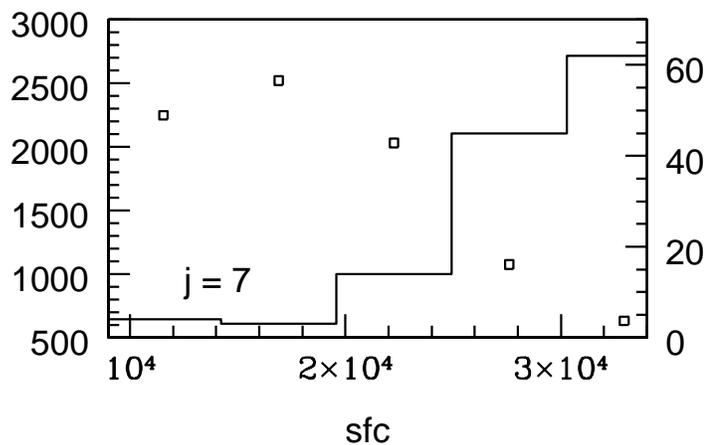
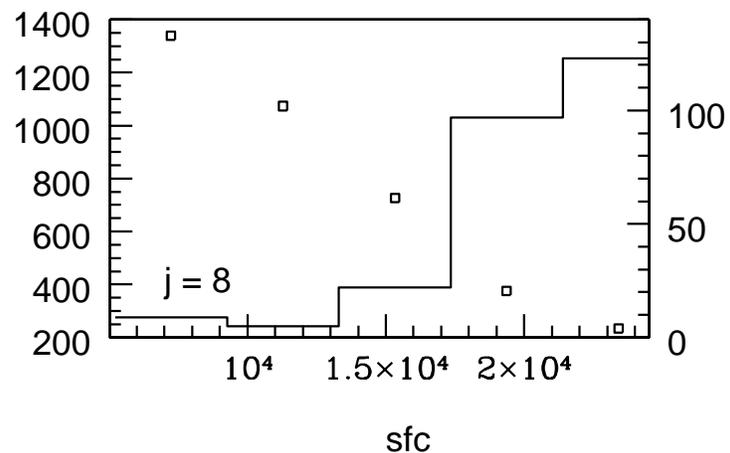
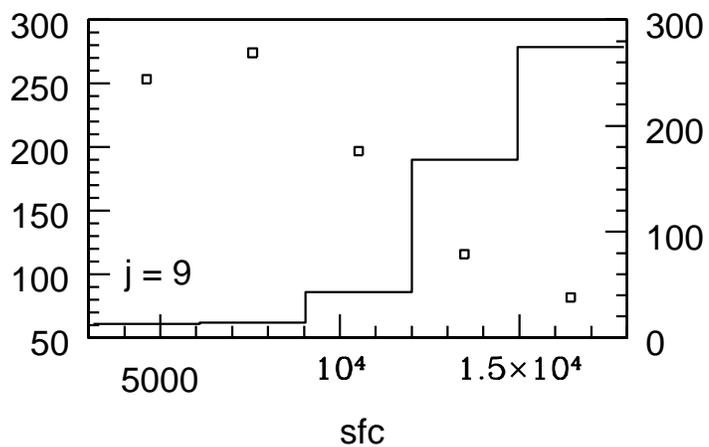
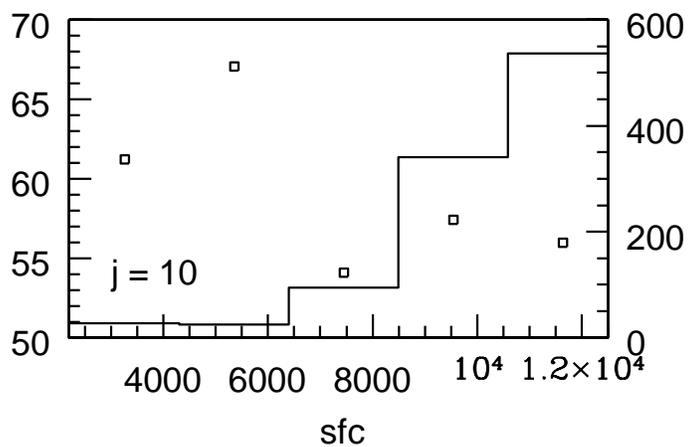
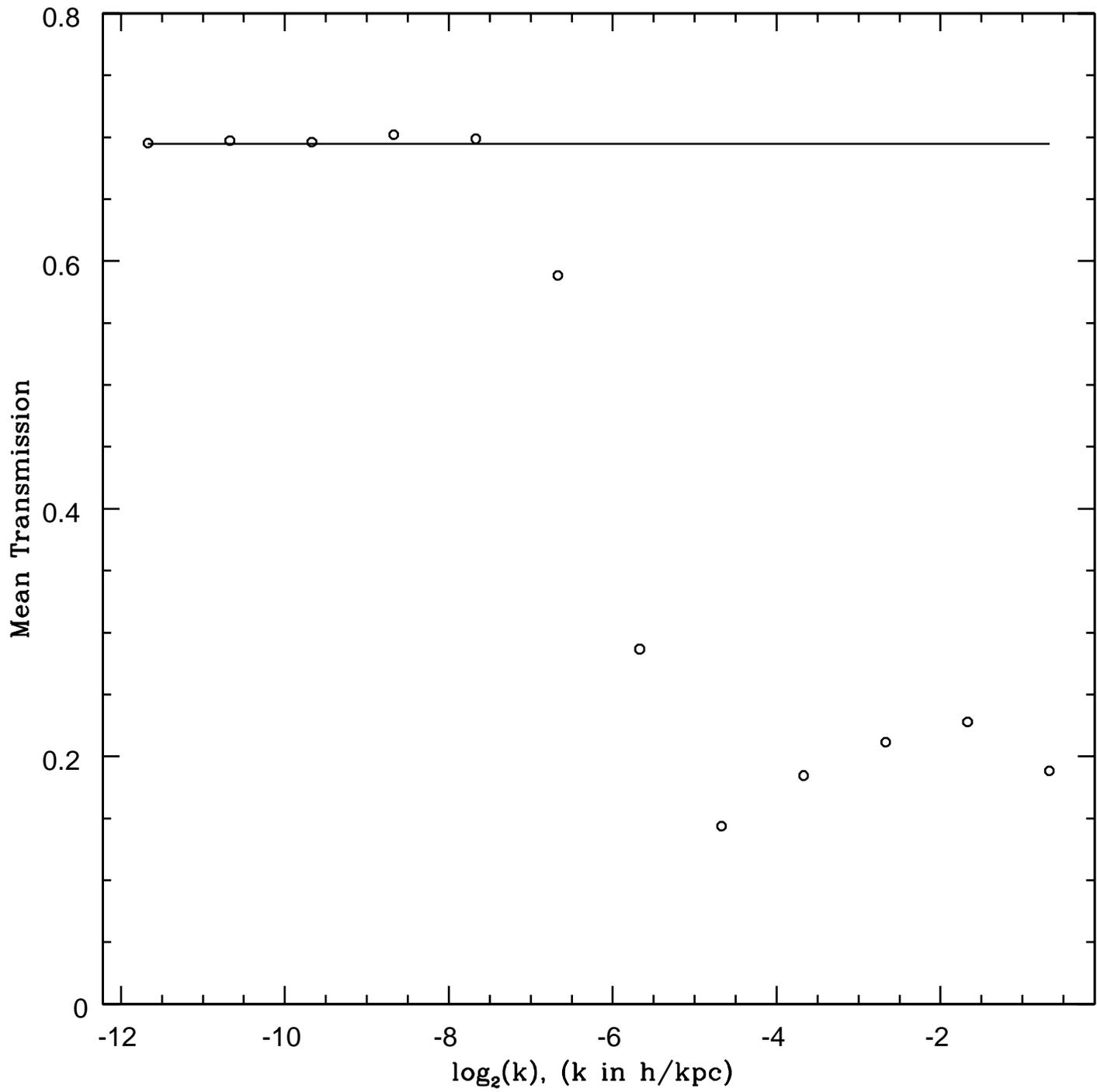
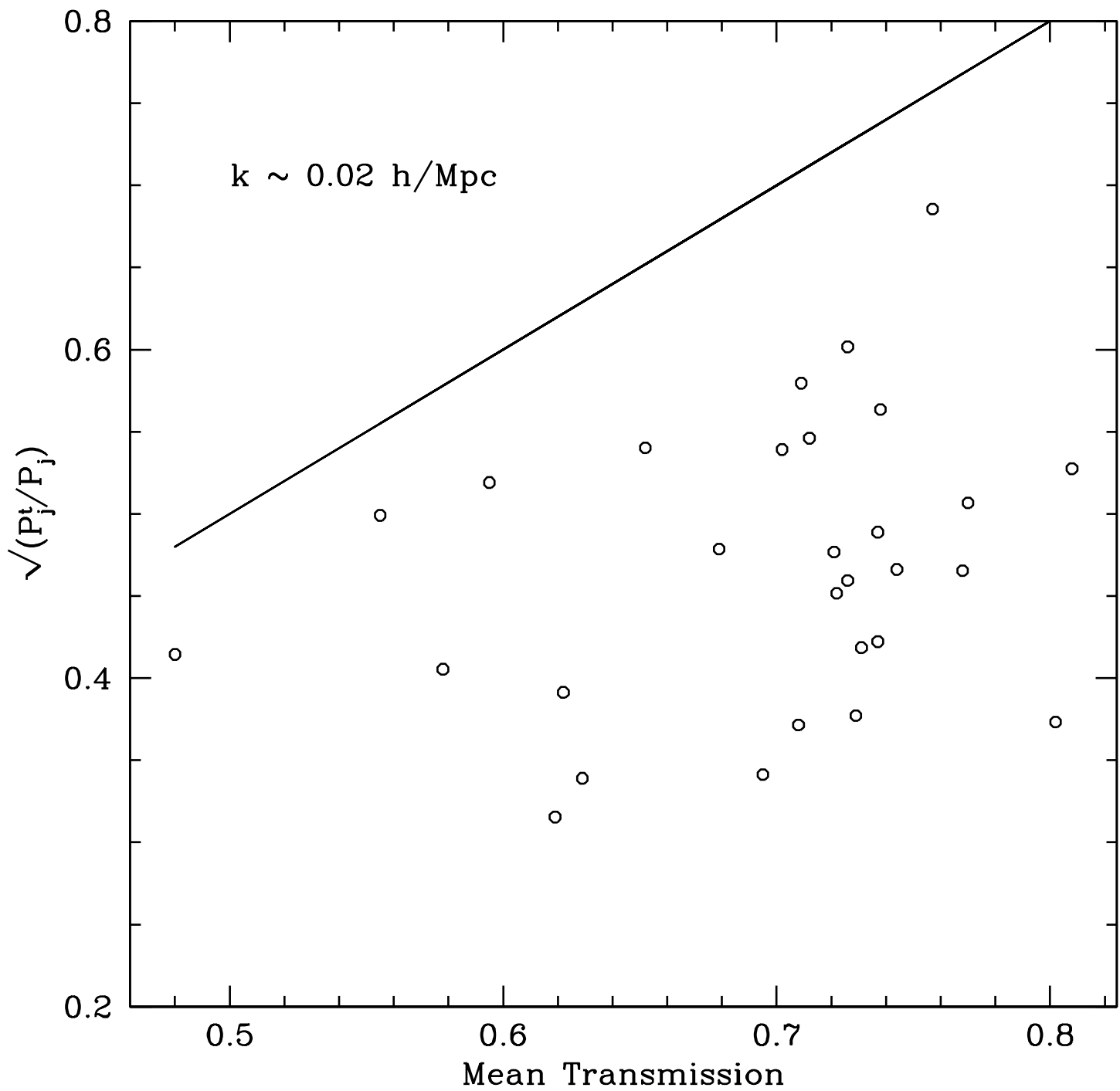
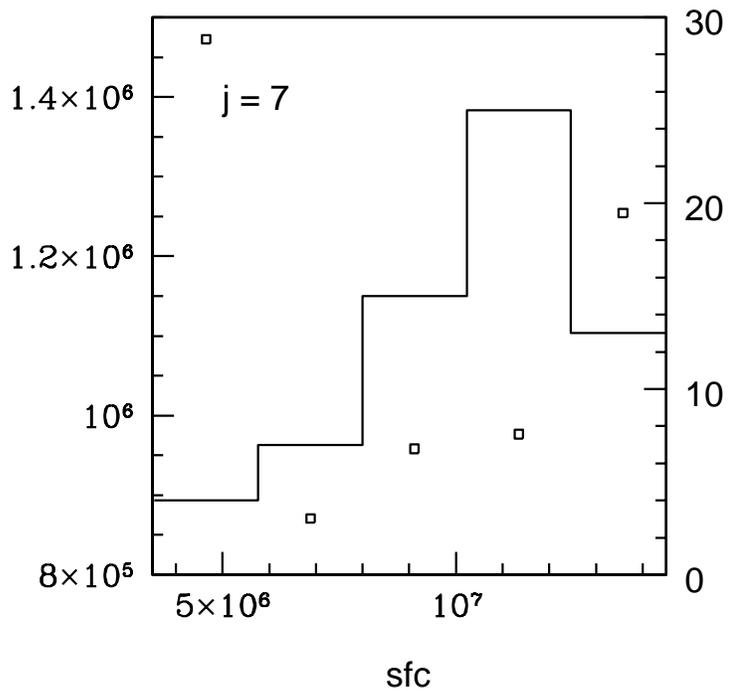
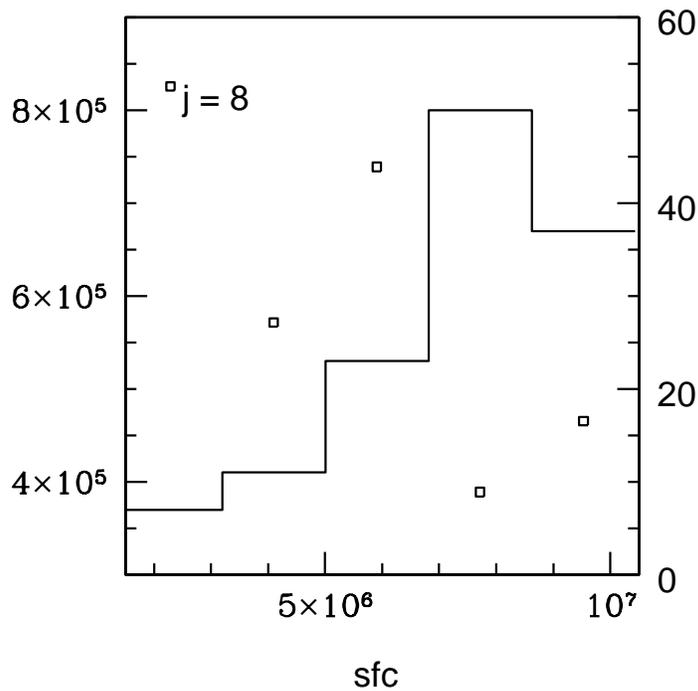
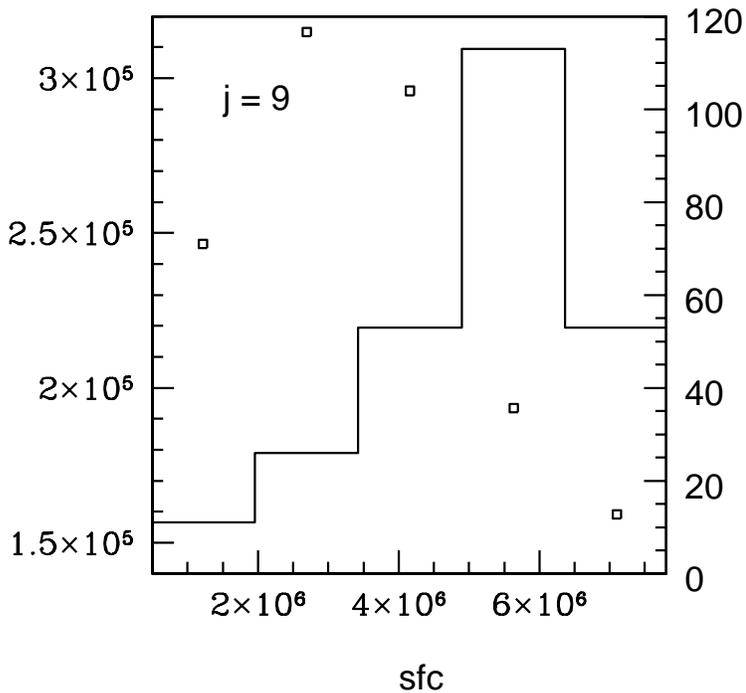
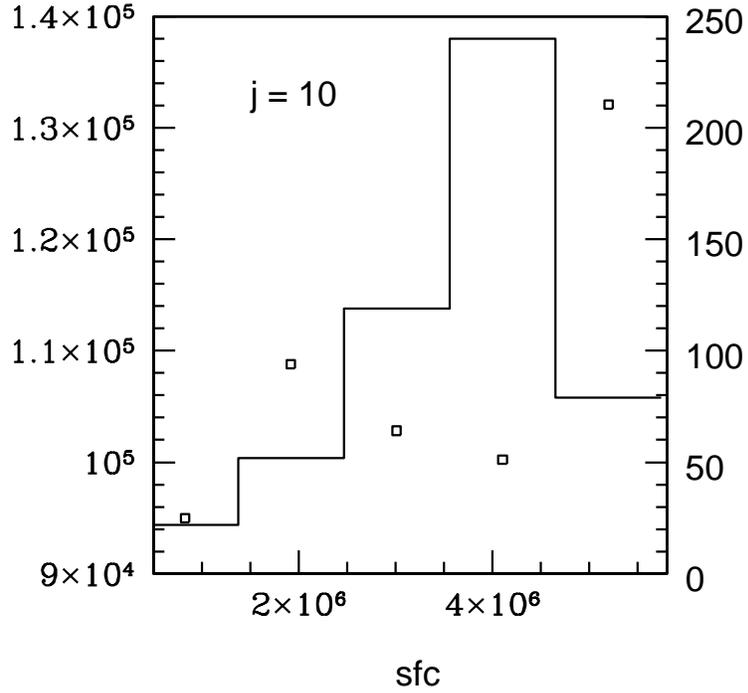
\end{document}